# A Cyber Infrastructure for the SKA Telescope Manager

Domingos Barbosa*[a], Joao Paulo Barraca[a,b], Bruno Carvalho[c] , Dalmiro Maia[d], Yashwant Gupta[e] , Swaminathan Natarajan[f] ,Gerhard Le Roux[g] , Paul Swart[g]
[a]Instituto de Telecomunicações, Campus Universitário de Santiago, 3810-193 Aveiro, Portugal;
[b]Universidade de Aveiro, Campus Universitário de Santiago, 3810-193 Aveiro, Portugal;
[c]Critical Software, Parque do Taveiro, Lote 49 3045-504, Coimbra, Portugal;
[d]CICGE, Faculdade de Ciências da Universidade do Porto, R. do Campo Alegre, Porto, Portugal;
[e]NCRA-TIFR, Pune University Campus, Post Bag 3, Ganeshkhind Pune 411007, India;
[f]TCS Research, Tata Consultancy Services, IITM Research park, Chennai 600 113, India;
[g]SKA South Africa, 3rd Floor, The Park, Park Road, Pinelands, 7405, South Africa

**ABSTRACT**

The Square Kilometre Array Telescope Manager (SKA TM) will be responsible for assisting the SKA Operations and Observation Management, carrying out System diagnosis and collecting Monitoring & Control data from the SKA sub-systems and components. To provide adequate compute resources, scalability, operation continuity and high availability, as well as strict Quality of Service, the TM cyber-infrastructure (embodied in the Local Infrastructure - LINFRA) consists of COTS hardware and infrastructural software (for example: server monitoring software, host operating system, virtualization software, device firmware), providing a specially tailored Infrastructure as a Service (IaaS) and Platform as a Service (PaaS) solution. The TM infrastructure provides services in the form of computational power, software defined networking, power, storage abstractions, and high level, state of the art IaaS and PaaS management interfaces. This cyber platform will be tailored to each of the two SKA Phase 1 telescopes (SKA_MID in South Africa and SKA_LOW in Australia) instances, each presenting different computational and storage infrastructures and conditioned by location.  This cyber platform will provide a compute model enabling TM to manage the deployment and execution of its multiple components (observation scheduler, proposal submission tools, M&C components, Forensic tools and several Databases, etc). In this sense, the TM LINFRA is primarily focused towards the provision of isolated instances, mostly resorting to virtualization technologies, while defaulting to bare hardware if specifically required due to performance, security, availability, or other requirement.

**Keywords:** Radioastronomy, SKA, infrastructure, virtualization technologies, Cloud Computing

## 1. INTRODUCTION

The Square Kilometre Array (SKA) is an international multipurpose next-generation radio interferometer, an Information and Communication Technology machine with thousands of antennas linked together to provide a collecting area of one square kilometer [1]. The SKA is the only global project in the European Strategy Forum of Research Infrastructures (ESFRI), with 10 Full members (Australia, Canada, China, Germany, Italy, New Zealand, South Africa, Sweden, The Netherlands and the United Kingdom) and an Associated member (India). It further involves more than 67 organizations in 20 countries, and counts with world-leading Information, Computing and Telecommunication (ICT) industrial partners. The SKA will be built by Phases in the Southern Hemisphere, first with Phase 1 in South Africa and Australia (starting in 2018), spreading in Phase 2 (starting in 2022) to SKA African Partners - Botswana, Ghana, Kenya, Zambia, Madagascar, Mauritius, Mozambique, Namibia - and Australia/New Zealand. SKA will be a central core of ~200 km diameter, with 3 spiral arms of cables connecting nodes of antennas spreading over sparse territories in several countries up to 3000km distances.  Since the SKA will continuously scan the sky, it will present a strong need for Quality of Service (QoS) of its IT infrastructure to achieve high operational availability.

*dbarbosa@av.it.pt; phone +351 234 377 900; fax +351 234 700 901

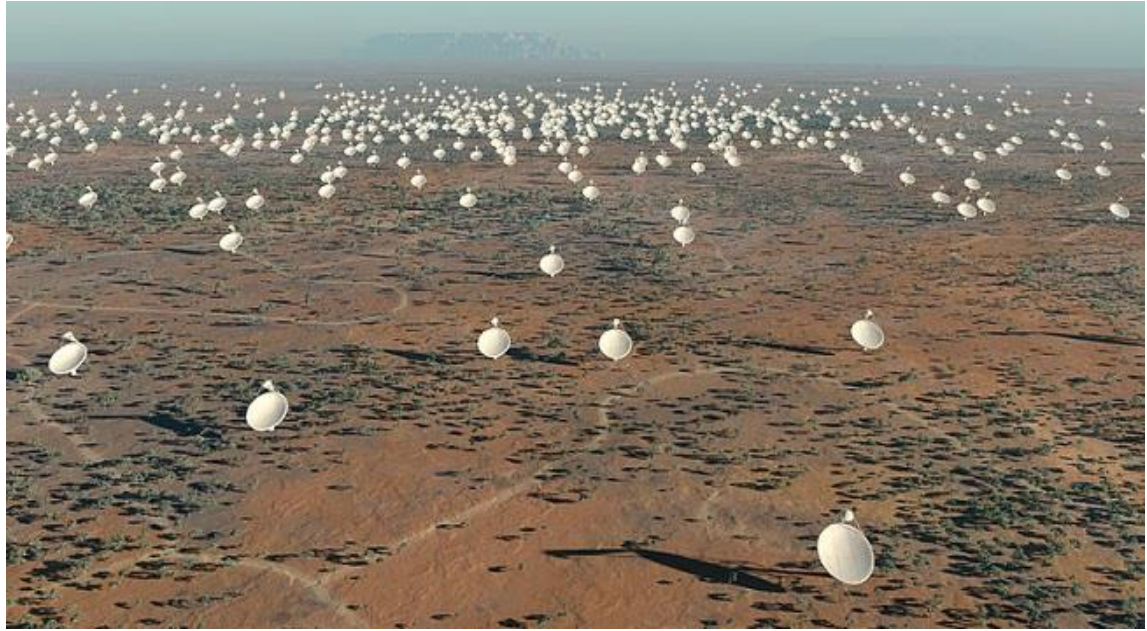

Figure 1 - An artist vision of the SKA Core site, with some of the projected 3000 15-meter parabolic dishes. From http://www.skatelescope.org.

The SKA is broken down in to various elements that will form the final SKA telescope managed by an international consortium comprising several world leading experts in their respective fields. We should note that SKA will be in fact constituted by two Telescopes, SKA_Low(Australia) and SKA_MID (South Africa), each one providing their own Telescope Manager (TM) subsystems [2]. Table 1 provides the characteristics of each SKA telescope. A distinct TM product will be designed for each Telescope during the Pre-Construction Phase. These TM products will be based on a common architecture and design as far as possible.

|  | **SKA_LOW (Australia)** | **SKA_MID (South Africa)** |
|---|---|---|
| **Sensors type** | 130 000 dipoles | 197 Dishes (including 64 MeerKAT) |
| **Frequency range** | 50-350 MHz | 0.45-15 GHz |
| **Collecting Area** | 0.4 Km$^2$ | 32 000m$^2$ |
| **Max baseline** | 65 Km (between stations) | 150 Km |
| **Raw Data Output** | 157 Terabyte /sec<br>0.49 Zettabyte/year | 3.9 Terabyte/sec<br>122 Exabyte/year |
| **Science Archive** | 0.4 Petabyte/day (128 Petabyte/year) | 3 Petabyte /day (1.1 Exabyte/year) |

Table 1 - Broad Characteristics of the two SKA Phase 1 Telescopes

## 2. THE SKA TELESCOPE MANAGER

The TM is a computer-based system that integrates Telescope equipment, monitors it for health and status, exercises high-level control over (coordinating) it as required to perform observations, and provides a human interface for the Telescope. The TM will be responsible for the Observation Management, carrying out System diagnosis and collecting Monitoring & Control data from the other SKA Elements [3]. The TM high level functions are allocated to the TM sub-elements: the Observation Management (OBSMGT), the Telescope Management (TELMGT), its own Local Monitoring & Control (LMC), and its local Infrastructure (LINFRA),ie the TM compute platform.

The TM has three core responsibilities:

- Assisting the user in management of astronomical observations;

- Assisting the user in management of the telescope hardware and software sub-systems in order to perform those astronomical observations;

- Management of the data to support operators, maintainers, engineers and science users in achieving operational, maintenance and engineering goals; this excludes management of the science data products (such as visibilities, images, catalogues), which are the responsibility of Science Data Processor (SDP) Element.

To support these responsibilities, the TM performs high-level functions, namely: proposal handling, observation management, telescope management and data management. To support proposal handling, the TM provides tools for proposal generation and submission by the science user, and for the evaluation, assessment and approval of and time allocation for the proposals by the proposal evaluation committee.

While performing observation management, the TM allows scientists or Principal Investigators (PIs) to generate Program Blocks, which consist of Scheduling Blocks (SBs). The SBs contain information needed to schedule and execute observations for a Proposal. The SBs are then scheduled for execution. As observations are made, the TM provides the current telescope configuration and dynamic status to the SDP element for annotating the matching scientific data or visibilities.

While executing observations and performing telescope management, the TM orchestrates the appropriate elements, collects monitor data that is used to track the status of all Elements of the Telescope, and responds to faults, threats and opportunities. The TM manages monitoring and configuration data as a system model that describes the status of the telescope at any one time. The TM continually collects telescope configuration, telescope dynamic status and environmental data. The data is time-stamped and stored. The TM provides the data to users as the current and historic state of the system to support operations and maintenance. It maintains a version-controlled Telescope Model describing the current processing applied in the Telescope and distributes as required.

TM includes a self-monitoring and control function that configures the rest of TM, monitors it's functioning, performs fault management as needed, and provides lifecycle support functionality including upgrades and shutdown. The presence of this independent function contributes greatly to the reliability of TM. Dependability is also a major consideration in the design of the Telescope Manager infrastructure, including redundancy and failover both for computation and communications.

We stress the Telescope Manager is the agent of operations for the System, and as such, is very dependent on the SKA Operational Concept. Many of its underlying design assumptions are connected to the operational aspects described in the SKA Technical Use Cases [4]. We stress that while the processes and timelines for telescope commissioning and roll-out are being addressed and refined, the TM has a key role to play in supporting the commissioning and verification of other Elements. As such, the SKA TM is inherently dependent on the design details of other Elements for some key operational requirements, because of its role as a service provider to other Elements. These interface requirements between TM and the other Elements are defined via the Interface Control Document (ICD) process, that provides clarity and additional emergent requirements.

The TM functional layers can be described as:

- User interface: This layer gives the user access to the functions of the TM.

- High level Telescope operations: Both OBSMGT and TELMGT implement some higher level functions, for instance, proposal handling, observation scheduling and execution, alarm monitoring, etc. This layer provides content for user interfaces as well as an API for observation scripts.

- Utility layer: TELMGT provides lower level services that are used by OBSMGT and also by some TELMGT functions. Typical functions in this layer are: acquisition of monitoring data from LMCs of other Elements, aggregation of monitoring data, the translation of science oriented commands to Element level commands. The use of aggregation nodes enables parallelization and scalability. The TM LMC supports this role as the local monitoring and control, including support to lifecycle management. This layer provides abstraction of data and access to control over Telescope technology.

- Local infrastructure: This layer is provided by LINFRA and consists of the mechanical and electrical components that interface with the SKA Infrastructure (INFRA) element), networking equipment that interface with the SKA Signal and Digital Transport (SADT) element and the computing infrastructure for TELMGT, OBSMGT and LMC products. In practice, the TM element Local Infrastructure (TM. LINFRA) is a sub-element that consists of COTS hardware and infrastructural software (for example: server monitoring software, host operating system, virtualization software, device firmware and other necessary middleware software for Quality of Service assurance and support to the other subelements layers).

The TM articulated and synchronized relations to the other elements for each telescope, like the SKA_MID DISH, MeerKAT Dishes, SKA_LOW Low Frequency Aperture Array (LFAA), SDP, Central Signal Processing (CSP), SADT and INFRA can be perceived from the following **Figure 2**:

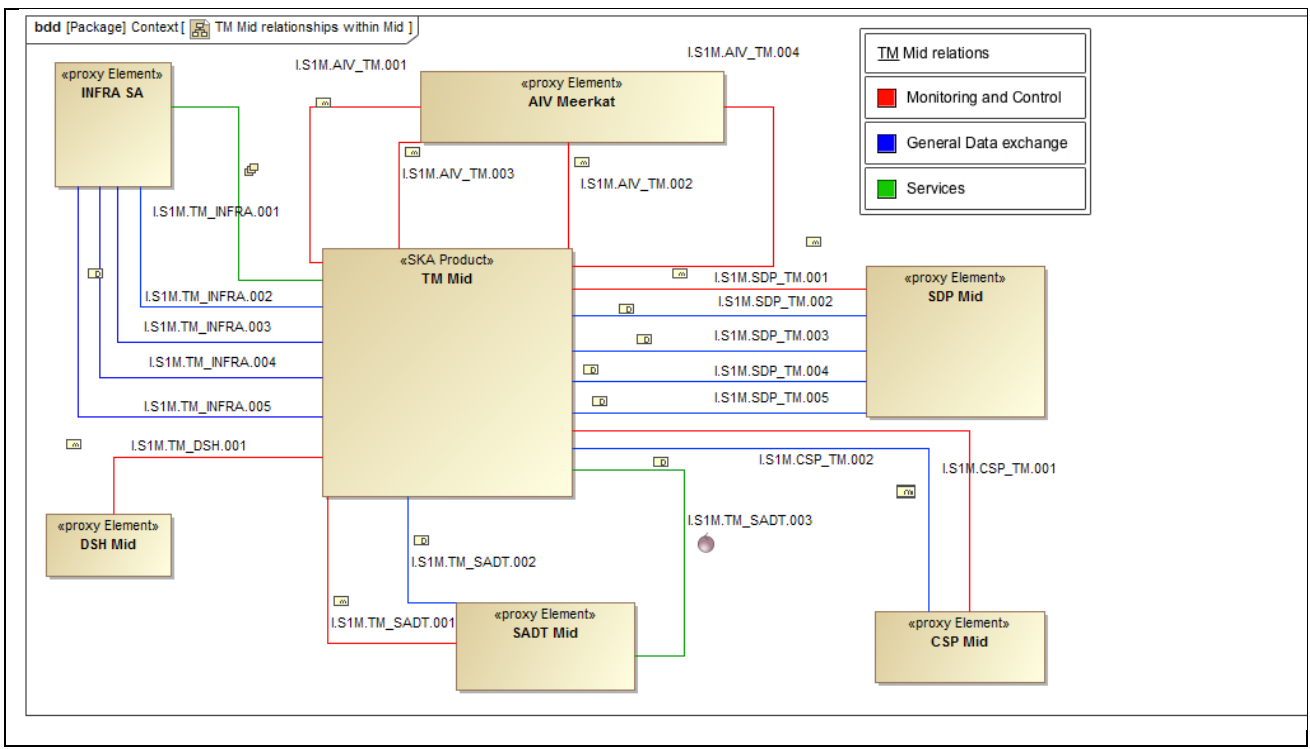

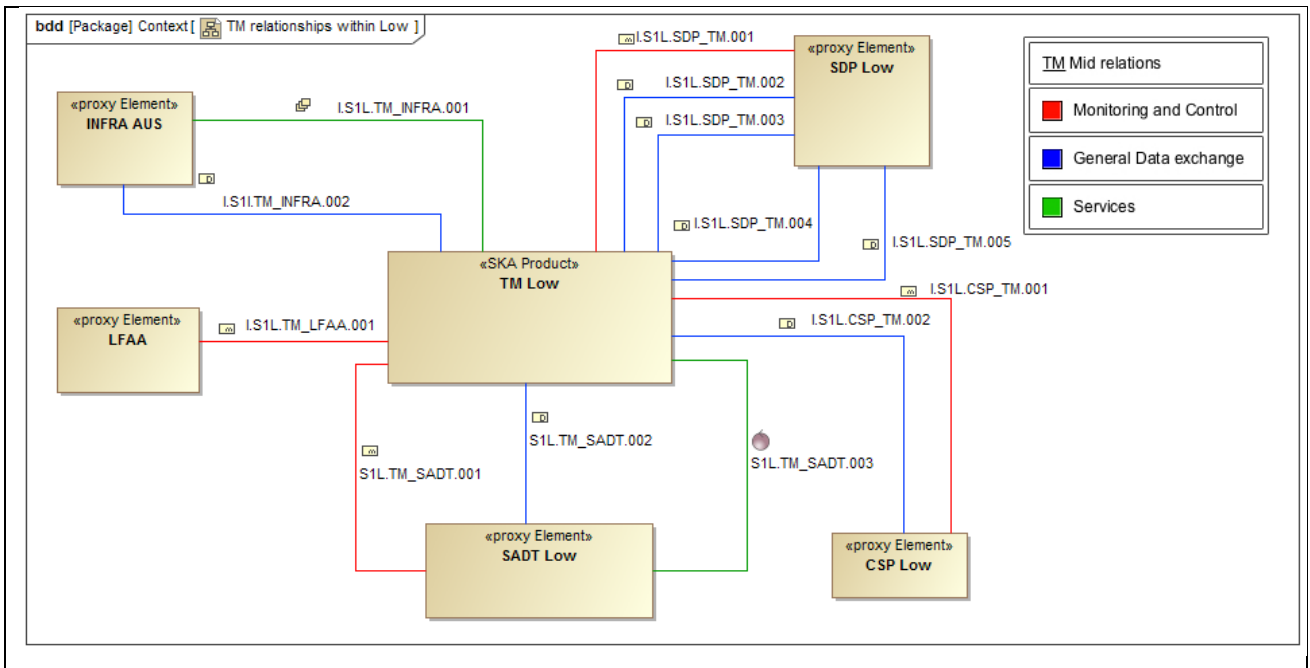

Figure 2 - TM Relations to the other SKA Elements. Top: TM to SKA_MID relations; Bottom: TM to SKA_LOW relations.

## 3. THE TELESCOPE MANAGER LOCAL INFRASTRUCTURE

The TM LINFRA provides the following services to the other TM sub-elements, including its LMC : Computational, networking, power and facilities infrastructure for each TM sub-element and component. LINFRA is a sub-element that consists of COTS hardware and infrastructural software (for example: server monitoring software, host operating system, virtualisation software, device firmware, etc. LINFRA will consist primarily of COTS hardware (servers, storage and networking interfaces / switches) [5, 7]. Software will consist mainly of Operating System software, server monitoring software, support software and device firmware. LINFRA will vary accordingly with the telescopes, presenting different computational and storage infrastructures, as can be perceived from the Deployment Concept model (**Figure 3**), but sharing a high degree of similarity of the physical and electrical interfaces between TM LINFRA and INFRA across the two telescopes. Likewise, the lower layers of networking interfaces between TM LINFRA and the SADT Element will be similar for both telescopes. It is considered that LINFRA will expose an Infrastructure as a Service (IaaS) operation model and for some TM products may expose a Platform as a Service (PaaS), hosting the other software layer's products, therefore removing management complexity from specific service oriented products that require a high degree of availability.

These TM products will differ for each telescope and will deployed at the different core sites: the bulk of TM will be co-located inside the Central Processing Facilities (CPFs), one in the Karoo Central Astronomy Advantage Area (SA) and the other in the Murchison Radio Observatory (AUS), with instances and monitoring/operating controls in facilities nearby in the Karoo (Klerefontein) and Cape Town (in SA), and in Boolardy and Perth (AUS). Furthermore, TM will have a footprint at the Observatory HQ, where proposal handling and operations snapshots may be executed.

Under this model, LINFRA is primarily focused towards the provision of isolated instances, mostly resorting to virtualization technologies, while defaulting to bare hardware if specifically required due to performance, security, availability, or other requirement. TM executes sub-elements products in instances that can be deployed, migrated, replicated and upgraded, while providing key performance indicators regarding the execution conditions of each execution instance created:

- The Deployment functionality enables the easy reproducible deployment of instances from development into production environments.

- The Migration functionality enables the instances to be moved between physical hardware resources in order to increase availability and redundancy.

- The Replication functionality enables redundancy and automatic scaling to cope with bursts of requests.

- The Upgradability functionality enables components to be replaced by updated versions with fixes or new features.

For each instance, key operational parameters will be reported, allowing the SKA TM to track the correct execution condition of its components and maintaining high operational availability.

As an IaaS, LINFRA incorporates Cloud Computing Service tool technologies to provide specific Application Programming Interfaces (APIs) up to online services [6]. These will abstract the top software layers from the details of physical infrastructure and related aspects like computing resources, location, data partitioning, scaling, security, backup, etc.

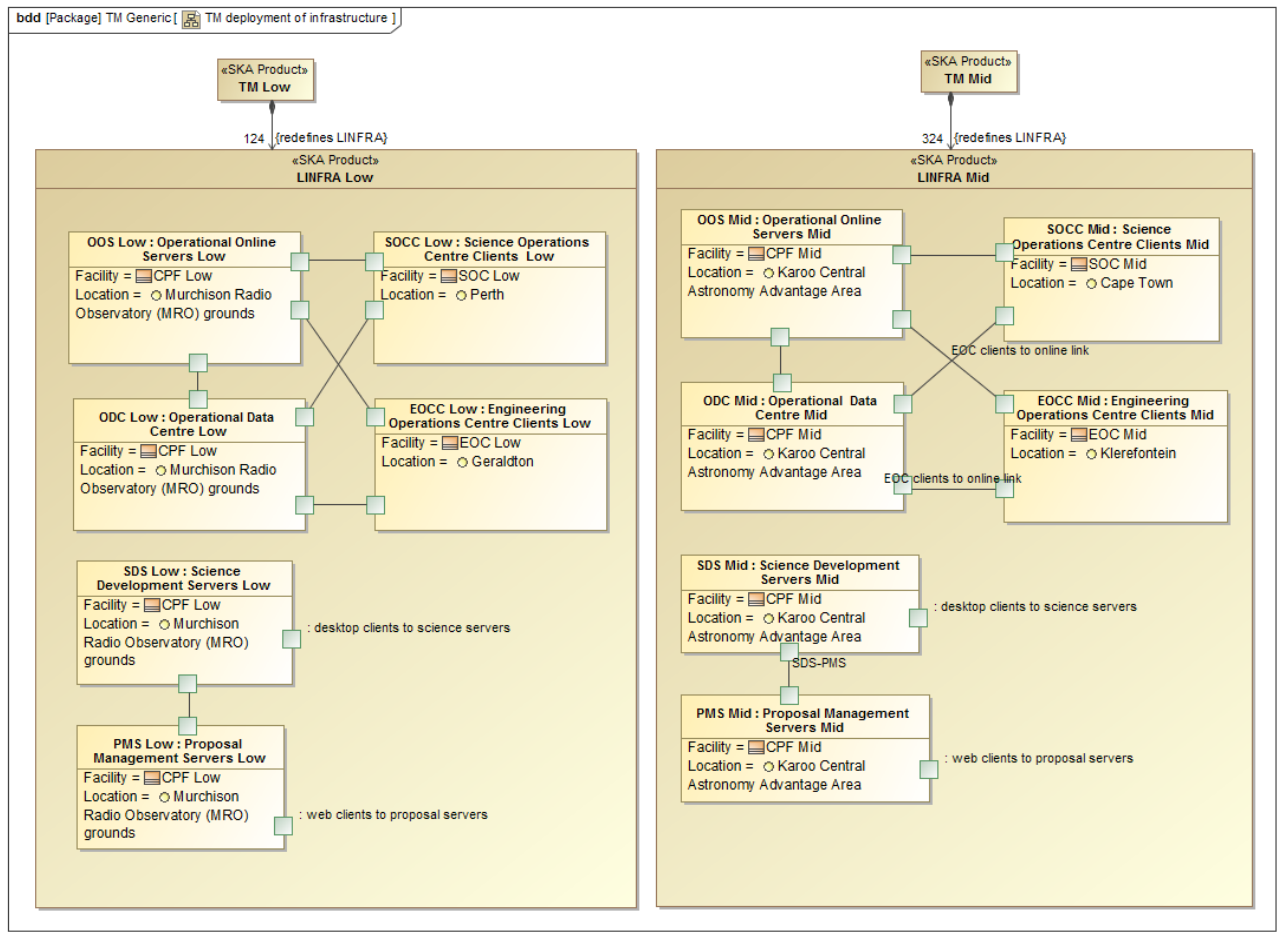

Figure 3 TM Infrastructure Deployment. Although most of the LINFRA will be co-located in the Central Processing Facility (CPF) of each telescope, several instances will also be deployed in the Science Operations Centres (SOC) and the Engineering Operation Centers (EOC).

The Main TM. LINFRA functions and Purposes:

- Interface with INFRA; Power distribution (as provided by INFRA to TM) to TM sub-elements.

- Support server operations and provide controlled hosting environment with basic monitoring & control functionalities (Provide computational infrastructure to OBSMGT, TELMGT, and LMC)

- Interface with SADT ; Network connectivity to co-located TM sub-elements

- Provide Support to TM Life cycle management/facilitation

- Provide capabilities to deploy and manage virtual instances

- Provide capabilities to migrate virtual instances due to power optimization, security, performance or failure avoidance.

- Provide the means to spawn instances at different locations using a distributed file system for data distribution.

The TM LINFRA will consist of a computational platform (hardware), virtualization orchestration software, and the operational network and power interfaces to other TM sub-elements whose properties and topologies will be determined by the data connections needs of sub element designs and the interface definitions set by INFRA and SADT, dependent on location. Furthermore, it will also manage a storage infrastructure for storing virtual images, snapshots and other data. Additional requirements traceability for both SKA1_MID and SKA1_LOW include Electromagnetic compatibility, Environmental conditions, Safety, Electrical power consumption.Specifically, LINFRA architecture can be divided in the components (Hardware, Software and Storage), described below.

The LINFRA hardware platform will consists of many dozens of individual compute nodes running one or more virtual instances. Each compute node will be mostly based on Commercial of the Shelf (COTS) equipment, with its corresponding networking, local storage subsystems, power distribution units and extensive cooling systems. Latency, bandwidth, computation requirements, memory and capacity will drive hardware dimensioning and location of each virtual instance. In particular, LINFRA hardware design will be such it minimises latency and reliability problems, and hence the probability of heavy message usage between servers and instances is low. Moreover, there should not exist any level of resource overcommit so that execution performance is highly predictable. Performance analysis for latencies, in particular for fault detection, will rely on a Failure Mode, Effects, and Criticality Analysis (FMECA) identification of faults. The TM.LINFRA electronic equipment will meet a commercially Electromagnetic Compatibility (EMC) standard, according for instance to EN 55022 EU standard  to ensure criteria compatibility between EU, SA and AUS, and compliance with Radio Spectral protection for the SKA Radio Quiet Zones (RQZs) according to the International Telecommunications Union (ITU) Radioastronomy band protection  recommendations (ITU-R RA.769-2). Overall, we may refer broad guidelines to be followed in the LINFRA sizing:

- Cost-efficient server hardware with support for virtualization (Intel VT-x, Intel EPT or other with similar performance)
- Large Symmetric multiprocessing (SMP) communication efficiency
- Large network communication efficiency
- Balanced Design.
- Interplay of storage and networking technology (disk locality is no longer relevant in intra-datacentre computations).
- Remote management interface such as the Intelligent Platform Management Interface (IPMI)

- High computation density and high memory density architectures
- Low power design, with support for scalable frequency and low power execution states.

Some systems that will need to be supported are:

- Redundant power supplies
- Redundant network cards

Networking will rely on 1-Gbps Ethernet switches with up to 48 ports as commodity components, to connect a single rack (including switch port, cable, and server NIC). Some of its systems will require higher performance, with one or more 10-Gbps interconnects inside the racks, and/or to other critical system, such as remote storage, high speed message buses, or an LMC. The cost/performance analysis will require a multi-level hierarchy for the internal network fabric.

The Software layer provided by LINFRA may be generically described as a platform providing orchestration and support of processing and storage services, which is typically considered an Infrastructure as a Service (IaaS) and Platform as a Service (PaaS) provider. Characteristics like ample parallelism, Workload churn, Platform homogeneity, Fault-free operation, presence of Service-level dashboards, performance debugging tools and platform-level health monitoring will be part of LINFRA software fabric to enable possible operation support to the other TM sub-elements At its lowest level there will be the software layer managing and monitoring hardware resources (servers, switches, racks) in order to maintain systems operational, minimize downtime, and predict evolution requirements.

Then there will be the common firmware, kernel, operating system distribution, interfaces, and libraries expected to be present in all individual servers. This requires the existence of tools for automating the deployment, configuration and upgrade of software components as required. This environment should allow the installation and configuration of servers and Operating Systems (OS), enabling the deployment of TM sub-elements in bare metal installations. Tools such as Chef and Puppet, which allow the automation of the installation process, and creation of homogeneously configured environments is desired.

On top of the hardware, there will be the hypervisor running virtualized environments capable of abstracting the hardware of a single machine and provide a custom execution environment. TM will default to deploying software components into virtual instances. Latency, computation, memory and performance requirements may impose the need for deploying software components into bare hardware.

Virtual instances are to be orchestrated through a specific orchestration middleware, capable of deploying, managing and migrating execution environments on demand and as required. Availability requirements also imposes the need for supporting the creation of snapshots in order to better store the execution state for later analysis. This orchestration middleware will expose programmatic and user interfaces allowing for remote management of virtual resources. This is vital for TM as it allows to maximize availability, better cope with upgrades to software elements, and better scale when needed.

LINFRA will also provide a generic block storage, available through the network fabric using standard protocols (e.g., NFS). This storage must be able to host any file, such as OS images, support software, and instance snapshots, as well as state information from TM sub-elements. The requirement is for TM to be able to store a history of the state of the system so that a past picture can be replayed to fidelity as close as possible to what is presently displayed. At this stage, the estimation of storage requirements can only be an estimate and stored data will vary according to monitored elements, being very fast for measurement sensors to very slow like alarms, status changes, etc. From current preliminary analysis, the SKA TM_MID will generate approximately 250TB of data required for storage over a 1-year period (without dissemination). The SKA TM_Low may require less than a 1-2 PB. Most likely, storage will be based on rotational technologies as part of a TM Storage Area Network (SAN) directly connected to the cluster-level switching fabric. Adoption of hierarchical storage will greatly contribute to reducing infrastructure costs associated.

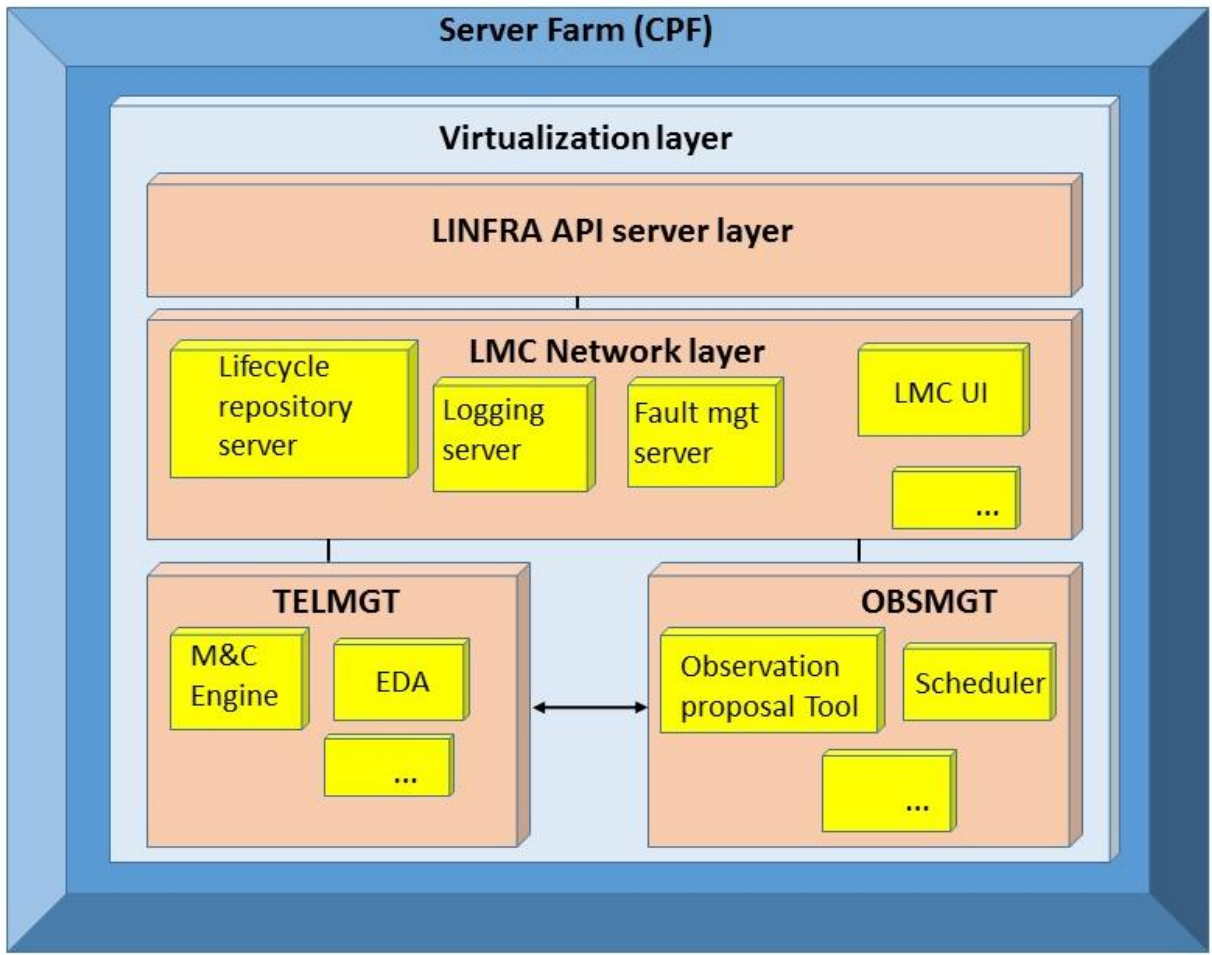

Figure 4 - TM LINFRA encapsulated layers (update for better fig/fonts)

LINFRA design also has to account for the TM Power consumption, an important aspect in remote zones like the SKA core sites where power budgets are limited. Indeed, power cap allocations may be imposed, forcing optimization of architectures. Power provision to TM will be a responsibility of the INFRA Element, including the routing of the Power Distribution Unit through the system UPS. Any threshold applying to TM is in fact being applied to LINFRA equipment deployment and operation. Extrapolation of peak power usage for IT equipment from both SKA Precursors (MeerKAT and ASKAP, MWA) have been studied under external interfaces to INFRA. However, we stress linear scalability is certainly not adequate, since on one hand virtualization techniques and compute and storage topologies may reduce power needs, and on the other hands technology roadmaps (ie Moore's law) indicate continuous server miniaturization towards more capacity with higher power efficiency. Thus, virtualization technologies existing at LINFRA can be used to reduce power consumption by migrating or freezing instances. This may apply to some functions not required to continuously run on the TM (e.g. high multiplicity of sub arrays, forensic analysis, low utilisation of dish and LFAA stations during observations). However, TM sub-elements should never be migrated or frozen while experiments, calibration or diagnostic tasks are running. LINFRA will provide the means to manage virtual instances and TM sub-elements will be responsible for stopping/freezing/migrating instances as adequate for reducing power consumption.

**3.1 LINFRA Technology Stack**

In order to address the risk mitigation objectives, the primary objectives of the LINFRA architecture activities can be decomposed in terms of : software, lifecycle, control and management architectures. In short, and considering SKA Availability requirements for down-time, LINFRA will deploy a high level of Automatisms, focus on core state of the art technologies; include for cloud APIs since end users are nowadays cloud aware implying culture changes for legacy application coding and IT services. TM.LINFRA shall be such it removes complexity for service deployment. The LINFRA Component Lifecycle itself is defined according to Products requirements, Service Catalogue, Key Performance Indicators (KPIs) and Dimensioning.

The main components can be perceived from the following table:

| | |
|---|---|
| **Instantiation:** | • Component requirement definition, SLA, dependencies, failover actions, monitoring parameters, ACLs, keys, network rules (centralized, versioning?)<br>• Matching to computation, storage and network resources. This requires Coordination with SADT for network<br>• Provisioning of configurations and software components Automation of software installation and configuration<br>• Instantiation of component container<br>• Identify Risks : which Application of configuration, networking rules, acls, etc...<br>• Availability of monitoring and control interface through API. E.g.: monitor component KPIs, issue configuration updates |
| **Monitoring:** | • Define quotas per service (SLA); Plan capacity upgrade; Detect malfunction. Risks : policy and degree of automatisms<br>• What to do when quota is exceeded? Trigger alarm to LMC<br>• Include Autoscale Bursting to public clouds.<br>• Risks : E.g. Proposal Submission Tool; Run some tools in public clouds? E.g.:dev env, forensics, other. |
| **Accounting** | • Ceilometer component (OpenStack) provides accounting; Track usage of each experiment; Better cost plan of experiments<br>• Capability of defining dynamic rules to trigger alarms |
| **Active** | • Security incidents; old: reinstall, new: replace with new VM<br>• Misconfiguration requiring reboot; Watchdogs, alarms, backup VMs (or hot standby)<br>• Resize a service; add VMs to serve demand of specific server<br>• Horizontal scalability; resize VMs (or rather, replace with bigger); Resize services automatically<br>• Under command of LMC or with information from SLA<br>• Risk: evaluate tools and borderline configuration |

Furthermore, from the several orchestration middleware options (Openstack, OpenNebula and VMWare ESX) and as part of a risk management, LINFRA is evaluating tools from the Cloud Computing Openstack ecosystem as part of its prototyping activities [6]. These tools were identified for being Open Source (no "enterprise" version), Open Design (Open design summit, can be customized), Open Development, and have an Open community (ie, access to wide support, examples, information availability) and are developed as a suite offering compatibility between each other and with core software technologies. The following components will be addressed: Dashboard, Identity Management, Networking, Load balancers, Database.

Current prototyping activities are using the following technologies:

- Linux based host systems

- Linux or Windows based virtual machine guest systems

- Virtualization components based on Hardware Virtual Machines, Para-virtualization, or software containers (e.g using Docker).

- OpenStack platform as a basis for virtualization platform.

- Technologies related to configuration tracking such as Chef, Puppet and Ansible .

- MySQL, Oracle and PostgreSQL databases, together with alternative solutions such as Cassandra and MongoDB. These solutions are also a subject of TELMGT detailed investigation as part of the Database solution for the Tango framework for LINFRA to host.

## 4. CONCLUSIONS

The Square Kilometre Array Telescope Manager (SKA TM) will be responsible for assisting the SKA Operations and Observation Management, carrying out System diagnosis and collecting Monitoring & Control data from the SKA sub-systems and components. Essentially, TM appears as Network peer to compute and storage, where the TM LINFRA provides an Infrastructure as a Service (IaaS) and Platform as a Service (PaaS), therefore removing complexity by providing automatisms to APIs software layers. The chosen tools must also be compatible with other managerial tools.

## 5. ACKNOWLEDGMENTS

DB and JPB acknowledge support from FCT through national funds and when applicable co-funded by FEDER – PT2020 partnership agreement under the project UID/EEA/50008/2013. DM acknowledge support from FCUP. The Portuguese team contributed through the ENGAGE SKA Consortium.